\documentclass[smallabstract,smallcaptions]{dccpaper}

\pdfoutput=1

\usepackage{epsfig}
\usepackage{amsmath}
\usepackage{amssymb}
\usepackage{color}
\usepackage{url}
\usepackage{cite}

\newlength{\figurewidth}
\newlength{\smallfigurewidth}

\def\bfx{{ \bf x  }}

\def\defeq{{\stackrel{\Delta}{=}}}
\def\calX{{\mathcal{X}}}

\newtheorem{theorem}{Theorem}[section]
\newtheorem{lemma}[theorem]{Lemma}
\newtheorem{proposition}[theorem]{Proposition}

\newenvironment{proof}[1][Proof]{\begin{trivlist}
\item[\hskip \labelsep {\bfseries #1}]}{\end{trivlist}}

\begin{document}

\title
{\Large
\textbf{Cutset Width and Spacing for\\ Reduced Cutset Coding of Markov Random Fields}
}


\author{%
{\small\begin{minipage}{\linewidth}\begin{center}
\begin{tabular}{ccc}
{Matthew G. Reyes}$^{\ast}$ & \hspace{20mm} & {David L. Neuhoff}$^{\dag}$\\
$^{\ast}$self-employed & \hspace*{0.5in} & $^{\dag}$EECS Dept., University of Michigan \\
\url{mgreyes@umich.edu} && \url{neuhoff@umich.edu}
\end{tabular}
\end{center}\end{minipage}}
 \thanks{An abbreviated version of this paper has been submitted
to ISIT 2016.\vspace{1ex}}
}


\maketitle
\thispagestyle{empty}


\begin{abstract}
In this paper we explore tradeoffs, regarding coding performance, between the thickness and spacing of the cutset used in Reduced Cutset Coding (RCC) of a Markov random field image model \cite{reyes2010}. Considering MRF models on a square lattice of sites, we show that under a stationarity condition, increasing the thickness of the cutset reduces coding rate for the cutset, increasing the spacing between components of the cutset increases the coding rate of the non-cutset pixels, though the coding rate of the latter is always strictly less than that of the former. We show that the redundancy of RCC can be decomposed into two terms, a correlation redundancy due to coding the components of the cutset independently, and a distribution redundancy due to coding the cutset as a reduced MRF. We provide analysis of these two sources of redundancy. We present results from numerical simulations with a homogeneous Ising model that bear out the analytical results. We also present a consistent estimation algorithm for the moment-matching reduced MRF for the cutset $U$.
\end{abstract}

\section{Introduction}\label{sec:introduction}
A Markov random field (MRF) $X = \{X_i : i \in V \}$ is a collection of random variables on an undirected graph graph $G=(V,E)$, where the nodes\footnote{We use the terms \emph{nodes}, \emph{sites} and \emph{pixels} interchangeably.} in $V$ are the random variable indices and the edges in $E$ represent direct dependencies between the random variables \cite{wain:03b}, and is often proposed as a model for many sources of data,
such as images. A family of MRFs on a graph $G$ is defined by a vector statistic $t$ having a component for each edge and each node.  An individual MRF within this family is indicated by an exponential parameter vector $\theta$ whose components correspond to the components of $t$. Since there has been relatively little development of algorithms or theory for the compression of MRFs \cite{anas:82, kont:03, reyes2009t, reyes2009b, reyes2010, reyes2011, reyes2014}, we feel that this is an important problem to consider. In this paper we explore design tradeoffs of the lossless Reduced Cutset Coding method introduced in \cite{reyes2010}.

Reduced Cutset Coding (RCC) is a two-stage algorithm for lossless compression of an MRF defined on an intractable graph, where tractability is with respect to Belief Propagation (BP) \cite{wain:03b,reyes2010,reyes2011}. The method consists, first, of suboptimal lossless encoding of a cutset $U\subset V$, chosen such that the subgraphs $G_U$ and $G_W$ induced by $U$ and $W\!\!=\!\!V\setminus U$, respectively, are tractable. The components of $X_U$ are encoded with Arithmetic Coding (AC) using BP to compute a {\em reduced MRF} coding distribution. A reduced MRF for $X_U$ is an MRF on the subgraph $G_U$ induced by $U$, with the statistic $t$ limited to $U$, and a possibly different exponential parameter vector $\tilde \theta_U$. Secondly, conditioned on the encoded cutset $X_U$, the component subsets of the remaining variables $X_W$ are encoded conditioned on their respective boundaries, again using AC, with BP used to compute the true conditional coding distributions of the variables in $X_W$ with
respect to the original MRF.

The rate of this scheme can be expressed as
\begin{eqnarray}
                 R & = & \frac{\mid U\mid}{\mid V\mid}R_U + \frac{\mid W\mid}{\mid V\mid}R_{W},\label{eq:rcc_ratedecomp}
\end{eqnarray}

\noindent where $R_U$ is the rate in bits per pixel for the cutset $U$, and likewise $R_{W}$ for the remainder $W$. Because $G_W$ is tractable for BP, the conditional coding distributions for the components of $X_W$ can be exactly computed. Thus AC will encode each component on average at its conditional entropy plus an overhead of one or two bits \cite{whitten87}. Since we have in mind the components of $W$ having many pixels, the rate $R_{W}$ is well-approximated by $ {1 \over |W|} H(X_W|X_U)$, the ideal coding rate for $X_W$ given $X_U$. Similarly, since $U$ is tractable for BP, the reduced MRF coding distribution can be computed exactly, and $R_U$ is well-approximated by the (normalized) cross entropy ${1 \over |U|} H(X_U \| \tilde X_U)$ between the marginal distribution for $X_U$ and the reduced MRF distribution for the same variables, which we denote  $\tilde X_U$, and which equals the entropy $ {1 \over |U|}  H(X_U)$ of $X_U$ plus the divergence ${1 \over |U|}  D(X_U||\tilde X_U)$ between the true and reduced MRF distributions for $X_U$.

It follows that the rate of this scheme exceeds the rate of an optimal code, which is
\begin{equation}
 {1 \over |V|}  H(X_U, X_W) = {|U| \over |V|} {1 \over |U|} H(X_U) + {|W| \over |V|} {1 \over |W|} H(X_W|X_U),
\nonumber
\end{equation}
by the divergence ${1 \over |V|}  D(X_U||\tilde X_U)$.
For a given cutset $U$, this divergence is minimized by choosing the parameter
vector $\tilde \theta_U$ to be that which causes the mean of the statistic $t_U$ of the reduced MRF $\tilde X_U$ to be the same as the mean of $t_U$ on the marginal $X_U$
of the original MRF $X$ \cite{amar:00,wain:03b,reyes2010}.
This is called the \emph{moment-matching parameter} and denoted $\theta_U^*$. In Section \ref{sec:mom_match} we present a consistent algorithm for estimating $\theta^*_U$ for a tractable subset $U$, and as such, for the rest of this paper we let $\tilde X_U$ denote this moment-matching reduced MRF. Even when divergence is minimized, one normally expects ${1 \over |U|} H(X_U)$ to be larger than ${1 \over |W|} H(X_W|X_U)$.

In the present paper we consider an MRF on an $M\times N$ rectangular lattice of sites. The statistic $t$ as well as the parameter $\theta$ are both row-invariant, and the image height $M$ is assumed to be very large, so that the sequences of rows of the image are assumed to form a stationary process. The cutset $U$ consists of $k+1$ evenly spaced $n_L\times N$ rectangular regions $L_1,\ldots,L_{k+1}$, referred to as {\em lines}, so that the $k$ components of  $G_{W}$ are themselves $n_S\times N$ rectangular regions $S_1,\ldots,S_k$, referred to as {\em strips}. This is an extension of the RCC method of \cite{reyes2010}, \cite{reyes2011}, which restricted $n_L$ to be 1.%
\footnote{Even though now $n_L$ can be larger than one, we continue to use the nomenclature of {\em lines}.}
Here, $M=kn_S + (k+1)n_L$, so that lines and strips alternate, beginning with a line and ending with a line. This class of cutsets was chosen to simplify both the algorithm and the analysis. For example, the lines (strips) can be transformed into a simple chain graph by grouping the pixels in each column of a line (strip) into one superpixel. If $n_L$ and $n_S$ are both moderate, for instance at most 10, then BP can be used to perform exact inference efficiently.

An interesting question is how the cutset parameters $n_L$ and $n_S$ affect the individual rates $R_U$ and $R_{W}$ as well as the weightings of $R_U$ and $R_{W}$ by the respective sizes of $U$ and $W$. First, consider $R_U$. The lines of $U$ are encoded independently with the respective moment-matching reduced MRF coding distributions. From the stationarity assumption, these moment-matching reduced MRFs are the same for each line, and therefore
$$  R_U = {1 \over |U|} H(X_U || \tilde X_U  )
=  {1 \over n_L N}  H(X_L \| \tilde X_L) = {1 \over n_L N} H(X_L) + {1 \over n_L N} D(X_L \| \tilde X_L) ,$$
where  $L$ denotes a block of $n_L$ consecutive rows of the image, $X_L$ is the subset of the MRF on $L$, and $\tilde X_L$ is the same random variables with the moment-matching reduced MRF distribution.
Next, by the Markov property and stationarity,
$$ R_W = {1 \over |W|} H(X_W | X_U)  = {1 \over n_S N}  H(X_S | X_{\partial S}),$$
where $S$ denotes $n_S$ consecutive rows, $\partial S$ denotes
the boundary of $S$, and $X_S$ and $X_{\partial S}$ are the respective subsets of random variables on $S$ and $\partial S$.  Therefore, as a function of line and strip widths, the
\emph{per-row rate}\footnote{The overall rate is the
per-row rate divided by the row width $N$.  From now on, we mainly focus on per-row rate to simplify expressions, and use an overbar to indicate such.}
is
$$
    \bar R(n_L, n_S)  =  {(k + 1) n_L \over k n_S + (k+1)n_L } {1 \over n_L}  H(X_L \| \tilde X_L)
          +  {k n_S \over k n_S + (k+1) n_L} {1 \over n_S}  H(X_S | X_{\partial S})  .
$$
When $k$ is large, this is well approximated by
\begin{eqnarray}
     \bar R(n_L, n_S)  &\approx&  {n_L \over n_L + n_S} {1 \over n_L}  H(X_L \| \tilde X_L)
          +  {n_S \over n_L + n_S} {1 \over n_S}  H(X_S | X_{\partial S}) \nonumber \\
          &=&  {n_L \over n_L + n_S} {1 \over n_L}  \big( H(X_L) + D(X_L \| \tilde X_L) \big)
          +  {n_S \over n_L + n_S} {1 \over n_S}  H(X_S | X_{\partial S})  . \nonumber
\end{eqnarray}

Intuitively, as the cutset line width $n_L$ increases, $R_U$ decreases because both ${1 \over n_L}H(X_L)$ and the divergence  ${1 \over n_L} D(X_L ||\tilde X_L)$ would decrease. However, the fraction of sites ${n_L \over n_L + n_S}$ encoded at the larger $R_U$ rate increases.  Hence, there is a potential tradeoff between choosing $n_L$ to be large in order to reduce the cutset rate, and choosing $n_L$ to be small in order to reduce the fraction of sites in the cutset. Similarly, as $n_S$ increases, the fraction of pixels ${n_S \over n_L + n_S}$ encoded at the lower rate increases, but one intuitively expects $\bar R_W = { 1 \over n_S}H(X_S|X_{\partial S})$ to increase. Again, a potential tradeoff.

On the other hand, since the overall rate is $R(n_L, n_S) = {1 \over |V|} H(X_V)+ {1 \over |V|} D(X_U||\tilde X_U)$, we see that the divergence term
${1 \over |V|} D(X_U||\tilde X_U)$ is the redundancy of the code, and
one can therefore focus on what makes it small.
%
Letting $ \Delta(n_L,n_S) \defeq  {1 \over |V|} D(X_U||\tilde X_U)$ denote
the \emph{redundancy of the code}, we will show that the per-row
redundancy has the form
\begin{eqnarray}
    \bar  \Delta(n_L, n_S) &  = &  {(k + 1) n_L \over k n_S + (k+1)n_L } {1 \over n_L}  D(X_L \| \tilde X_L)
          +  {k n_S \over k n_S + (k+1) n_L} {1 \over n_S}  I(X_{L_i};X_{L_{i-1}}) \nonumber \\
                & \approx &  {n_L \over n_L + n_S} {1 \over n_L} D(X_L \| \tilde X_L)
          +  {n_S \over n_L + n_S} {1 \over n_S}  I(X_{L_i};X_{L_{i-1}}) \nonumber
\end{eqnarray}
where $I(X_{L_i};X_{L_{i-1}})$ is the mutual information between the random variables $X_{L_i}$ on a line and the random variables $X_{L_{i-1}}$ on the previous line.
Note that in the above formula for redundancy, which is entirely due to the encoding of the lines, the first term, which we call the \emph{distribution redundancy} is due to use of the reduce MRF coding distribution
on each line and the second term, which we call the \emph{correlation
redundancy} is due the fact that lines are coded
independently.
Note also that while
 the redundancy is entirely due to encoding of the lines, the
 correlation redundancy
depends on the strip width $n_S$.
Moreover, since there is no correlation redundancy in the encoding of the first line, it is appropriate to think of $I(X_{L_i};X_{L_{i-1}})$ as a penalty per strip.
From this viewpoint, one would expect that increasing $n_L$ reduces the divergence per cutset pixel ${1 \over n_L} D(X_L||\tilde X_L)$, but increases the fraction ${n_L \over n_L + n_S}$ of the image included in the cutset.  Hence, it is not clear what is the best value for $n_L$.  Similarly, one would expect that information $I(X_{L_i};X_{L_{i-1}})$ decreases in $n_S$, while the fraction of pixels ${n_S \over n_L + n_S}$ increases in $n_S$. Therefore, it is likewise not clear what $n_S$ should be.

%

The results of this paper are to show the following results, most of which have been conjectured above. Under the stationarity assumption, the coding rate $R^S_{n_S}$ of a strip increases with $n_S$, the coding rate $R^L_{n_L}$ of a line decreases with $n_L$ when the moment-matching reduced MRF is used to encode the lines, and  $R^S_{n_S} < R^L_{n_L}$ for all choices of $n_S$ and $n_L$. We also present a consistent estimation algorithm for the moment-matching parameter $\theta^*_{U}$. We show that the divergence $D(X_U||\tilde X_U)$, equivalently the redundancy, can be decomposed into a correlation redundancy due to encoding the lines independently and a distribution redundancy due to approximating the lines as reduced MRFs, and present analysis of these two sources of redundancy. Numerical simulations with an Ising model illustrate the propositions.

In the rest of this paper, Section \ref{sec:background} provides background on MRFs and lossless coding and Section \ref{sec:rcc} provides an overview Reduced Cutset Coding in the current setting. Section \ref{sec:mom_match} presents an estimation algorithm for $\theta^*_U$, Section \ref{sec:tradeoffs} establishes the anticipated tradeoffs between cutset thickness and spacing, and finally, Section \ref{sec:simulation} discusses numerical simulations with an Ising model.


\section{Background}\label{sec:background}
We introduce notation for lossless coding of MRFs.

\subsection{Graphs and Markov Random Fields}
A {\em path} in a graph $G=(V,E)$ is a sequence of nodes, each successive pair of nodes being joined by an edge in $E$.  A graph is said to be {\em connected} if every pair of nodes $i,j\in V$ can be joined by some path, and {\em disconnected} otherwise. For any $U \subset V$, its \emph{boundary} $\partial U$ is the set of nodes not in $U$ connected by an edge to a member of $U$. The subgraph $G_U = (U,E_U)$ {\em induced by} $U$ is the graph consisting of nodes and edges contained in $U$. Likewise, the subgraph $G_{V\setminus U} $ is obtained by removing $U$ and all edges incident to it from $G$.  If $G_{V \setminus U}$ is disconnected, each maximal connected subset of $G_{V \setminus U}$ is called a {\em component}, and $G_{V \setminus U}$ is simply the collection of the (disjoint) subgraphs induced by the respective components. A subset $U\subset V$ is called a {\em cutset} if $G_{V\setminus U}$ consists of more than one component.

A family of MRFs is specified by an alphabet $\cal X$ and a vector statistic $t=(t_i, i \in V;  t_{i,j}, \{i,j\} \in E)$ defined on the site values at individual nodes and the endpoints of edges.\footnote{Properly, this is a {\em pairwise} MRF. Generalizations to other MRFs are straightforward.} That is, for a given image $\bfx = \{x_i: i\in V\}$, the function
$t_{ij}:\calX\times\calX\longrightarrow\mathbb{R}$ determines the contribution of the pair $(x_i,x_j)$ to the probability of $\bfx$, and similarly for $t_i:\calX\longrightarrow\mathbb{R}$. We say that $X$ is an MRF based on $t$. The entire family of MRFs based on $t$ is generated by introducing an exponential parameter  vector $\theta=(\theta_i, i \in V; \theta_{ij}, \{i,j\} \in E)$ where for each node $i$, and neighbor $j\in\partial i$, $\theta_i$ and $\theta_{ij}$ scale the
sensitivity of the distribution $p(G;\bfx;\theta)$ to the functions $t_i$ and $t_{ij}$, respectively. Specifically, for an MRF $X$ on $G$ based on $t$ with exponential parameter $\theta$, configuration $\bfx$ has probability $p(G;\bfx;\theta)$  given by
\begin{eqnarray}
    p(G;\bfx;\theta)
                & = & \exp\{\langle\theta,t(\bfx)\rangle - \Phi(\theta)\},\label{eq:mrf_1}
\end{eqnarray}
where  $\langle ~ , ~ \rangle$ denotes inner product, $\Phi(\theta)$ is the \emph{log-partition function}, and the arguments of $p(\cdot;\cdot;\cdot)$ indicate, respectively, the graph on which the MRF is defined, the configuration in question, and the exponential parameter on the graph. For a given exponential coordinate vector $\theta$, we let $\mu=\mu(\theta)$ denote the expected value of the statistic $t$ under the MRF induced by $\theta$, and we refer to $\mu$ as the {\em moment} of the MRF. The MRF distribution over all configurations is denoted $p(G;X;\theta)$, and the entropy of an MRF is denoted $H(G;X;\theta)$.

The conditional probability of a configuration $\bfx_W$ on subset $W\subset V$ given the values $\bfx_{U}$ on another subset $U\subset V$ is denoted $p(G;\bfx_W| \bfx_{U};\theta)$. It is straightforward to check that $p(G;\bfx_W| \bfx_{\partial W};\theta) = p(G;\bfx_W| \bfx_{V\setminus W};\theta)$ for all $W$, $\bfx_W$, and $\bfx_{\partial W}$. This is the {\em Markov Property}. The conditional distributions of random subfield $X_W$ given a specific configuration $\bfx_{\partial W}$, or on the random subfield $X_{\partial W}$, are denoted $p(G;X_W| \bfx_{\partial W};\theta)$ and $p(G;X_W| X_{\partial W};\theta)$, respectively. Likewise, $H(G;X_W | \bfx_{\partial W};\theta)$ and $H(G;X_W | X_{\partial W};\theta)$ are the respective conditional entropies of $X_W$ given a specific configuration $\bfx_{\partial W}$ or the random subfield $X_{\partial W}$.

For subset $U$, the marginal probability distribution on $X_U$ is denoted $p(G;X_U;\theta)$, where $p(G;\bfx_U;\theta)$ denotes the marginal probability of configuration $\bfx_U$. The {\em reduced MRF} distribution for $X_U$ on $G_U$ based on statistic $t_U$ with exponential parameter $\tilde\theta_U$ is denoted $p(G_U;X_U;\tilde\theta_U)$ and has the same form as in (\ref{eq:mrf_1}), where $\Phi_U(\tilde\theta_U)$ denotes the log-partition function for the reduced MRF. Similarly, $p(G_U;\bfx_U;\tilde\theta_U)$ denotes the probability of configurations $\bfx_U$ under the reduced MRF distribution. The statistic $t_U$ is inherited from the original statistic $t$. The marginal entropy of $X_U$ is denoted $H(G;X_U;\theta)$ while the entropy of a reduced MRF $p(G_U;X_U;\tilde\theta_U)$ is denoted $H(G_U;X_U;\tilde\theta_U)$.

\subsection{Belief Propagation and Lossless Coding}\label{sec:lossless}

In general, ones uses Belief Propagation (BP) \cite{wain:03b} to compute $p(G;\bfx_U;\theta)$ for a configuration $\bfx_U$. Since the inner product $\langle t_U(\bfx_U),\theta_U \rangle$ can be computed directly, BP is used to compute the log-partition function $\Phi(\theta)$, and more generally, to marginalize over $X_{V\setminus U}$. If $G$ has no cycles, then $p(G;\bfx_U;\theta)$ can be computed with complexity linear in the number of nodes in $V$. If $G$ has cycles, one can compute $p(G;\bfx_U;\theta)$ by grouping subsets of $V$ into supernodes such that the new graph is acyclic \cite{wain:03b}. In this case, complexity is exponential in the size of the largest supernode. A graph is said to be {\em tractable} if either $G$ has no cycles or if $G$ can be clustered into an acyclic graph where the size of the largest supernode is moderate. Similarly, a subset $U$ is said to be tractable if $G_U$ is tractable, in which case $p(G_U;\bfx_U;\tilde\theta_U)$ can be computed for  the reduced MRF  on $G_U$. Also, for tractable subset $W$, $p(G;\bfx_W | \bfx_{\partial W};\theta)$ can be computed for configurations $\bfx_W$ and $\bfx_{\partial W}$.

\begin{figure*}
    \centerline{    \hbox{
    \hspace{0.1in}
    \includegraphics[scale = .4]{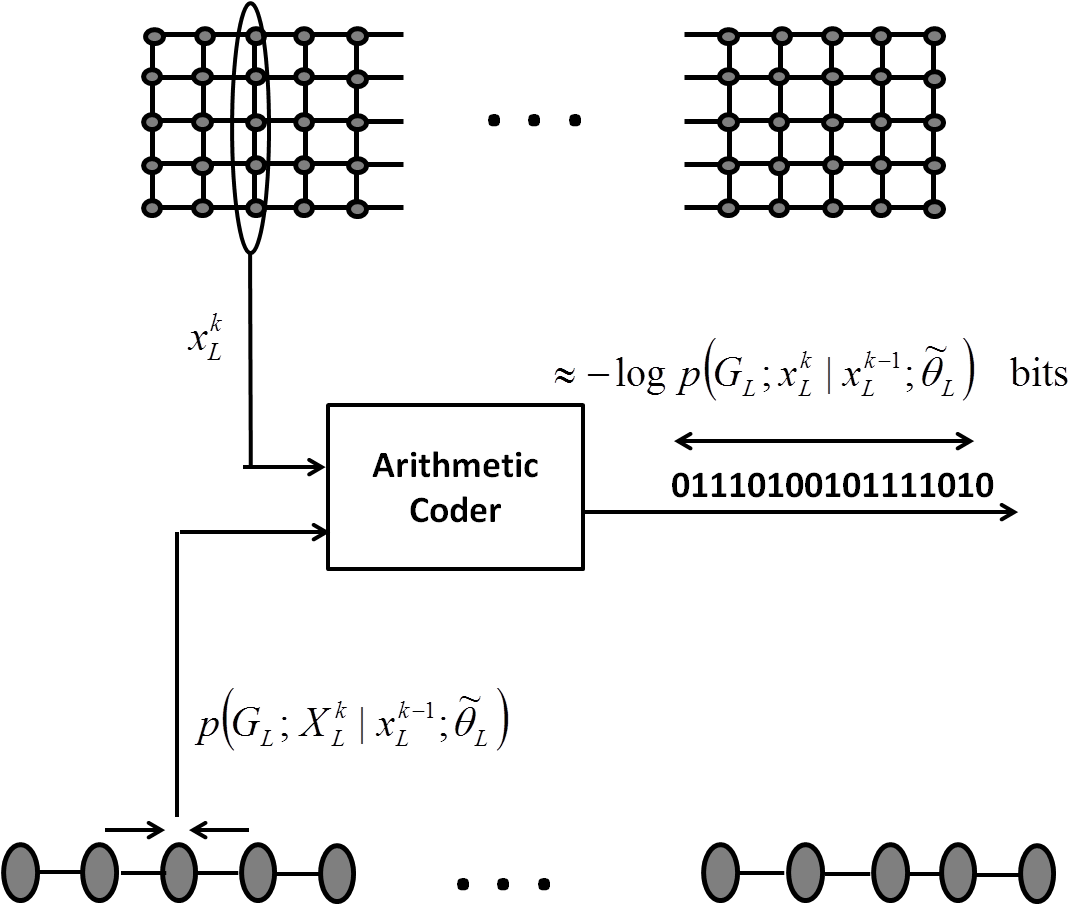}
    \hspace{0.2in}
    \includegraphics[scale = .4]{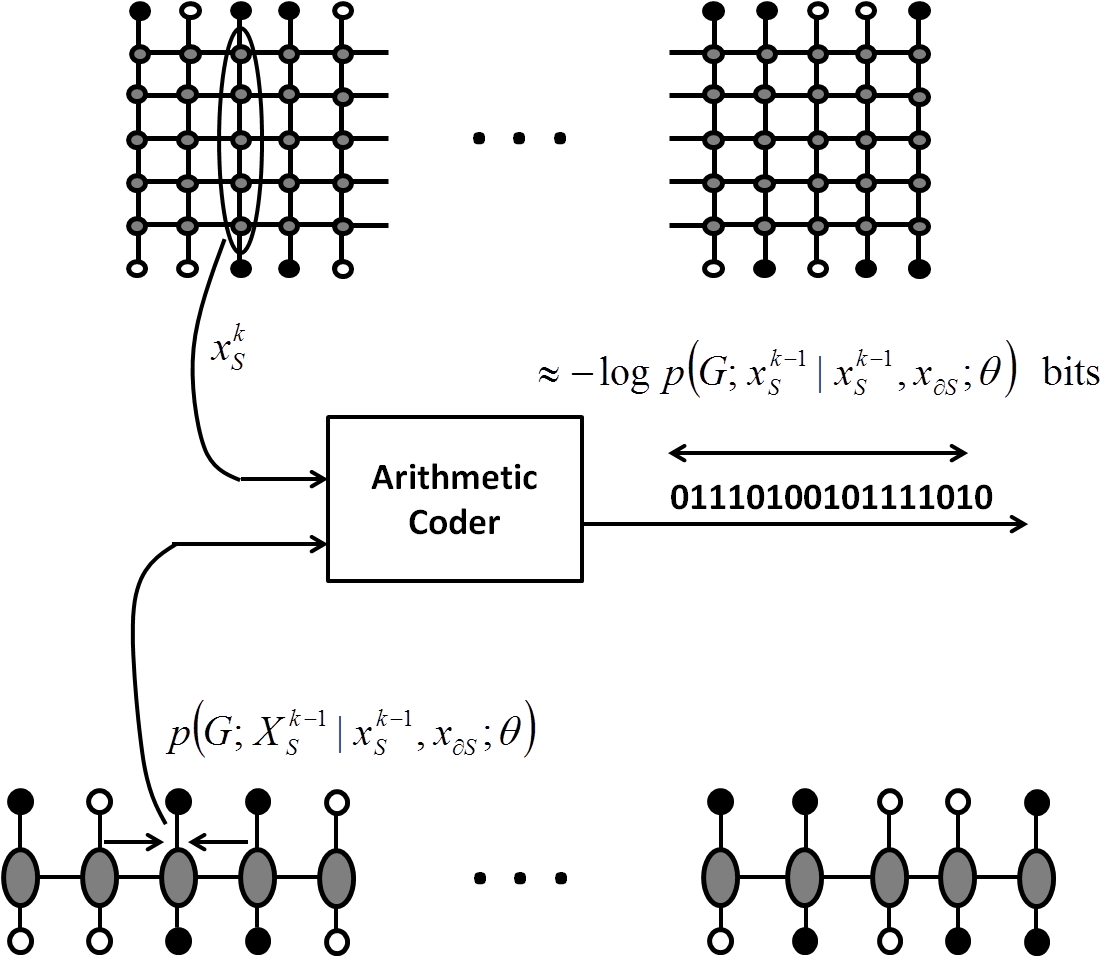}
    }   }
    \hbox{\hspace{1in} (a) \hspace{3in} (b)}
    \caption{(a) AC encoding of a line $X_L$ with reduced MRF $P(G_L;X_L;\tilde\theta_L)$ coding distribution using Belief Propagation, and (b) AC encoding of a strip $X_S$ conditioned on its boundary $X_{\partial S}$ with conditional distribution $p(G;X_S | X_{\partial S};\theta)$ using Belief Propagation.}
    \label{fig:coding}
\end{figure*}

For the purposes of this paper it suffices to say that lossless compression with an {\em optimal encoder} involves computation of a {\em coding distribution}. For a tractable subset $U$, if configuration $\bfx_U$ is losslessly compressed with reduced MRF coding distribution $p(G_U;X_U;\tilde\theta_U)$, then the average number of bits produced is the {\em cross entropy} $H(G;X_U;\theta || G_U;X_U;\tilde\theta_U)$ between the marginal distribution $p(G;X_U;\theta)$ and the reduced MRF coding distribution $p(G_U;X_U;\tilde\theta_U)$ for $X_U$, defined as
\begin{eqnarray}
    H(G;X_U;\theta || G_U;X_U;\tilde\theta_U) & = &
         H(G;X_U;\theta) + D(p(G;X_U;\theta) || p(G_U;X_U;\tilde\theta_U))\nonumber
\end{eqnarray}
\noindent where $D(p(G;X_U;\theta) || p(G_U;X_U;\tilde\theta_U))$ is the {\em divergence} from $p(G;X_U;\theta)$ to $p(G_U;X_U;\tilde\theta_U)$ and is the {\em redundancy} in the code \cite{cove:05}.

We showed in \cite{reyes2010} that the above divergence is minimized at $\theta^*_U$, the exponential parameter on $G_U$ such that the corresponding moment $\mu^*_U$ is equal to the moment subvector $\mu_U$ under the original MRF $p(G;X;\theta)$. The distribution of the reduced MRF $p(G_U;X_U;\theta^*_U)$ is called the {\em moment-matching} reduced MRF distribution for $X_U$, denoted $\tilde X_U$. When the moment-matching reduced MRF $p(G_U;X_U;\theta^*_U)$ is used as the coding distribution to encode $X_U$, the cross entropy is in fact the entropy $H(G_U;X_U;\theta^*_U)$ of the moment-matching reduced MRF \cite{reyes2010}.


For a tractable subset $W$, if configuration $\bfx_{W}$ is encoded conditioned on $\bfx_{\partial W}$ using coding distribution $p(G;X_W|\bfx_{\partial W};\theta)$, then the average number of bits produced is $H(G;X_W|X_{\partial W};\theta)$. Therefore, encoding $\bfx_W$ conditioned on $\bfx_{\partial W}$ is optimal, i.e., there is no redundancy.

In \cite{reyes2010}, Arithmetic Coding (AC) was proposed as the optimal encoder. Figure \ref{fig:coding} illustrates the encoding of a line and a strip. The mathematical details of using AC in the encoding of an MRF are given in \cite{reyes2009t}, \cite{reyes2010}, and \cite{reyes2011}, specifically in Chapter VI of \cite{reyes2011}.

\vspace{2mm}

\section{Reduced Cutset Coding}\label{sec:rcc}

In general, since the cutset $U$ consists of disjoint lines, the entropy of the moment-matching reduced MRF on $G_U$ is actually the sum $\sum_{L_i}H(G_{L_i};X_{L_i};\theta^*_{L_i})$ of the entropies of the reduced MRFs on the individual lines. Similarly, the conditional entropy of $X_W$ given $X_{\partial W}$ is the sum $\sum_{S_i}H(G;X_{S_i} | X_{\partial S_i};\theta)$ of the conditional entropies of the individuals strips given their respective boundaries.

In the present paper, we simplify this by considering vertically homogeneous parameters for the MRF, i.e., the components of the statistic $t$ and the exponential parameter $\theta$ do not vary vertically within the image. Furthermore, focusing only on the middle $M' = (k'+1) n_L + k' n_S \approx M/2$ rows of $V$, therefore excluding boundary effects, the image will be roughly stationary in the vertical direction. We let $B_n$ be an $n\times N$ rectangular subset of sites.

The random field $X_{B_{n_L}}$ on a line is encoded with reduced MRF coding distribution $p(G_{n_L};X_{B_{n_L}};\theta^*_{B_{n_L}})$. Normalizing by the number of pixels, the per-row rate for encoding a line is then
\begin{eqnarray}
   \bar R^L_{n_L} & = & \frac{1}{n_L }H(G;X_{B_{n_L}};\theta || G_{n_L};X_{B_{n_L}};\theta^*_{B_{n_L}}) \nonumber\\
             & = & \frac{1}{n_L }H(G_{B_{n_L}};X_{B_{n_L}};\theta^*_{B_{n_L}}). \nonumber
\end{eqnarray}

The random field $X_{B_{n_S}}$ on a strip is encoded conditioned on $X_{\partial B_{n_S}}$ with coding distribution $p(G;X_{B_{n_S}}|X_{\partial B_{n_S}};\theta)$. The per-row rate for encoding a strip is then
\begin{eqnarray}
    \bar R^S_{n_S} & = & \frac{1}{n_S }H(G;X_{B_{n_S}}\mid X_{\partial B_{n_S}};\theta). \nonumber
\end{eqnarray}

We let $\bar R_{n_S,n_L}$ denote the total per-row rate of RCC with cutset parameters $n_S$ and $n_L$, given by
\begin{eqnarray}  \label{eq:RSL}
    \bar R(n_S,n_L) \!\!\!\! & = & \!\!\!\! \frac{(k+1)n_L}{(k+1)n_L + kn_S} \bar R^{L}_{n_L} + \frac{kn_S}{(k+1)n_L + kn_S} \bar R^S_{n_S}. \nonumber
\end{eqnarray}
\noindent Assuming further that $M'$ is very large relative to $n_L$ and $n_S$, so that $k$ is very large, this rate is well-approximated by
\begin{eqnarray}  \label{eq:RSL}
    \bar R (n_S,n_L) & \approx & \frac{n_L}{n_L+n_S} \bar R^{L}_{n_L} + \frac{n_S}{n_L+n_S} \bar R^S_{n_S}. \label{eq:rate_approx}
\end{eqnarray}
\noindent

We now see that the performance of RCC with cutset parameters $n_S$ and $n_L$ is characterized by the rates $\bar R^L_{n_L}$ and $\bar R^S_{n_S}$, and the fractions $\frac{n_L}{n_L+n_S}$ and $\frac{n_S}{n_L+n_S}$.

\vspace{2mm}

\section{Moment-matching $\theta^*_U$}\label{sec:mom_match}

Recall from the previous section that the cross-entropy $H(p(G;X_U;\theta) || p(G_U;X_U;\tilde\theta_U)$ between the marginal distribution $p(G;X_U;\theta)$ of subset $X_U$ within an MRF on $G$ with statistic $t$ and a reduced MRF $p(G_U;X_U;\tilde\theta_U)$ on $G_U$ with statistic $t_U$ is minimized by the parameter $\theta^*_U$ such that the expected value $\mathbb{E}_{\theta^*_U}[t_U(X_U)]$ of the statistic $t_U$ in the reduced MRF equals the expected value $(\mathbb{E}_{\theta}[t(X)])_U$ of the statistic $t$ under the original MRF on the subset $U$, referred to as the {\em moment-matching} parameter. We will estimate $\theta^*_U$ from $n$ observations $\bfx^{(1)}_U,\ldots,\bfx^{(n)}_U$ on $U$, by seeking an $\hat\theta^n_U$ that minimizes an empirical version of the cross entropy, at least approximately. First, some background.

We let $\Theta = \{\theta\}$ denote the set of parameter vectors for MRFs on $G$ based on the statistic $t$. We restrict attention to the case where $\Theta$ is the subset of $\mathbb{R}^{|V| + |E|}$ of $\theta$'s with positive components.  In this case, due to the openness of $\Theta$, the family of MRFs based on $t$ is said to be {\em regular} \cite{wain:03b}. For parameter $\theta\in\Theta$, the function
\begin{eqnarray}
    \Lambda(\theta) & \defeq & \mathbb{E}_{\theta}[t(X)] \nonumber \\
                    & \defeq & \mu \nonumber
\end{eqnarray}
maps $\theta$ to $\mu$, the expected value of $t$ under the MRF induced by $\theta$, referred to as the {\em moment} of the MRF. The set $\mathcal{M} = \{\mu=\Lambda(\theta):\theta\in\Theta\}$ is the set of {\em achievable moments} for MRFs on $G$ based on $t$. We assume that the statistic $t$ is {\em minimal} in that the components of $t$ are affinely independent, meaning that the components of $t(\bfx)$ do not sum to a constant for all configurations $\bfx$. In this case, the function $\Lambda(\cdot)$ is one-to-one \cite{wain:03b}. Then, for $\mu\in\mathcal{M}$, the inverse function
\begin{eqnarray}
    \Lambda^{-1}(\mu) & = & \theta \nonumber
\end{eqnarray}
is well-defined. Moreover, $\mu$ is a dual parameter to $\theta$, in that the MRF $p(G;X;\theta)$ can alternatively be expressed as $p(G;X;\mu)$. For the MRF induced by parameter $\theta$, the subvector of moments on the set $U$ is given by
\begin{eqnarray}
    \Lambda_U(\theta) & = & \mu_U \nonumber
\end{eqnarray}
which can be seen as the restriction of $\Lambda(\cdot)$ to the set $U$.

For reduced MRFs on $G_U$ based on statistic $t_U$, $\tilde\Theta_U$ denotes the associated set of exponential parameters.
Now, consider the function
\begin{eqnarray}
    \tilde\Lambda_U(\tilde\theta_U) & = & \tilde\mu_U , \nonumber
\end{eqnarray}
which maps a parameter $\tilde\theta_U \in \tilde \Theta_U$ to the corresponding moment $\tilde\mu_U$ for the reduced MRF $p(G_U;X_U;\tilde\theta_U)$ on $G_U$. Likewise, $\tilde{\mathcal{M}}_U=\{\tilde\mu_U=\tilde\Lambda_U(\tilde\mu_U):\tilde\theta_U\in\tilde\Theta_U\}$ denotes the set of achievable moments for reduced MRFs on $G_U$. Since we have assumed that the statistic $t$ for the original family of MRFs on $G$ is minimal, the statistic $t_U$ for the family of reduced MRFs on $G_U$ is also minimal, and the inverse map $\tilde\Lambda_U^{-1}(\tilde\mu_U) = \tilde\theta_U$ is well-defined. Again, a reduced MRF $p(G_U;X_U;\tilde\theta_U)$ can also be parameterized as $p(G_U;X_U;\tilde\mu_U)$.


Given a parameter $\theta$ for an MRF $p(G;X;\theta)$, a subset $U$, and a sequence of observations $\bfx^{(1)}_U,\ldots,\bfx^{(n)}_U$ on $U$, we define the {\em empirical moment} of $p(G;X_U;\theta)$ as
\begin{eqnarray}
    \hat\mu^n_U & \defeq & \frac{1}{n}\sum\limits_{i=1}^nt_U(\bfx^{(i)}_U) . \nonumber
\end{eqnarray}
While $\mu_U=\Lambda_U(\theta)$ is always contained in $\tilde{\mathcal{M}}_U$, it is not necessarily the case that the empirical moment $\hat\mu^n_U$ is contained in $\tilde{\mathcal{M}}_U$. However, even if $\hat\mu^n_U$ is not in $\tilde{\mathcal{M}}_U$, $\hat\mu^n_U$ is still a limit point of $\tilde{\mathcal{M}}_U$ \cite{wain:03b}, meaning that for every $\epsilon > 0$, there is an $\epsilon$-ball containing $\hat\mu^n_U$ that contains infinitely many points of $\tilde{\mathcal{M}}_U$. Moreover, as stated in the following proposition, as the number of observations $n$ approaches $\infty$, not only is $\hat\mu^n_U$ in $\tilde{\mathcal{M}}_U$, but $\hat\mu^n_U$ converges to $\mu_U$.

\vspace{2mm}

\begin{proposition}\label{prop:momconv}
    The empirical moment $\hat\mu^n_U$ converges in probability to $\mu_U$, i.e.,
    for any $\epsilon >0$,
 \begin{eqnarray}  \label{eq:convergeinprob}
     \Pr \big( \big| \hat \mu^n_U - \mu_U \big| \leq \epsilon \big) \rightarrow 1, \mbox{ as } n \rightarrow \infty .
\end{eqnarray}
\end{proposition}

\vspace{2mm}

\begin{proof}
    To prove the proposition, one should recall that on a finite graph $G$, there does not exist a phase transition \cite{georgii}, and therefore, there is a unique MRF on $G$ for the specified statistic $t$ and exponential parameter $\theta$. It follows that the sequence $\bfx^{(1)}_U,\ldots,\bfx^{(n)}_U,\ldots$ is not only stationary but also ergodic, from which the proposition follows \cite{grimmett}. This completes the proof. $\hfill \Box$
\end{proof}

\vspace{2mm}

We now discuss the empirical version of cross entropy that we will minimize
as a surrogate for cross entropy. From a sequence of observations $\bfx^{(1)}_U,\ldots,\bfx^{(n)}_U$, we define the \emph{empirical cross entropy}
\begin{eqnarray}
    H_U^n(\hat\mu^n_U || \tilde\theta_U) & \defeq &-  {1 \over n} \sum\limits_{i=1}^n \log p(G_U;\bfx^{(i)}_U;\tilde\theta_U)  \nonumber \\
      &=&  -\sum_{ \bfx _U}   f( \bfx_U :  \bfx^{(1)}_U,\ldots,\bfx^{(n)}_U  ) \log p(G_U;\bfx_U;\tilde\theta_U) \nonumber
\end{eqnarray}
between the empirical distribution $f( \bfx_U :  \bfx^{(1)}_U,\ldots,\bfx^{(n)}_U  )$
generated by $ \bfx^{(1)}_U,\ldots,\bfx^{(n)}_U$  
and the reduced MRF $p(G_U;X_U;\tilde\theta_U)$ induced by a candidate parameter $\tilde \theta_U$.
That it makes sense to consider the empirical cross entropy to be a function of the
empirical moment $\hat \mu_U^n$ is due to the proposition presented later.
If $U$ is a tractable subset, then the probabilities in the summation can be efficiently computed.

Now, our estimate for the moment-matching parameter $\theta^*_U$ will be the $\hat \theta^n _U$ that minimizes this empirical cross entropy, at least approximately. It is well-known that $\Phi_U(\tilde\theta_U)$ is convex in $\tilde\theta_U$, and, as follows from the following theorem, so is the empirical cross-entropy $H^n_U(\hat\mu^n_U || \tilde\theta_U)$.  If, as we have assumed, the components of $t_U$ are affinely independent,  then $\Phi_U(\tilde\theta_U)$ and hence $H_U^n(\hat\mu^n_U || \tilde\theta_U)$ is strictly convex.  Therefore, either  $H^n_U(\hat\mu^n_U || \tilde\theta_U)$ has a unique minimum at a $\tilde\theta_U$  at which the gradient of  $H^n_U(\hat\mu^n_U || \tilde\theta_U)$ is zero, or since $\tilde\Theta_U$ is open,  $H^n_U(\hat\mu^n_U || \tilde\theta_U)$ does not have a minimum but approaches an infimum at a limit point of $\tilde\Theta_U$.  Moreover, from the following theorem and the fact that for any $\hat \mu^n_U$  there exists $\tilde \theta_U$ such that $\tilde \Lambda_U(\tilde \theta_U)$ is arbitrarily close to $\hat \mu^n_U$, we can find $\tilde \theta_U$ such that  the gradient is arbitrarily small and such $\tilde \theta_U$ must come  arbitrarily close to attaining the infimum of $H_U^n(\hat\mu^n_U || \tilde\theta_U)$. In either case, our ``moment-matching" estimate $\hat\theta^n_U$ will be a $\tilde \theta_U$ that induces a very small gradient.

\begin{figure*}
    \centerline{    \hbox{
    \hspace{0in}
    \includegraphics[scale = .5]{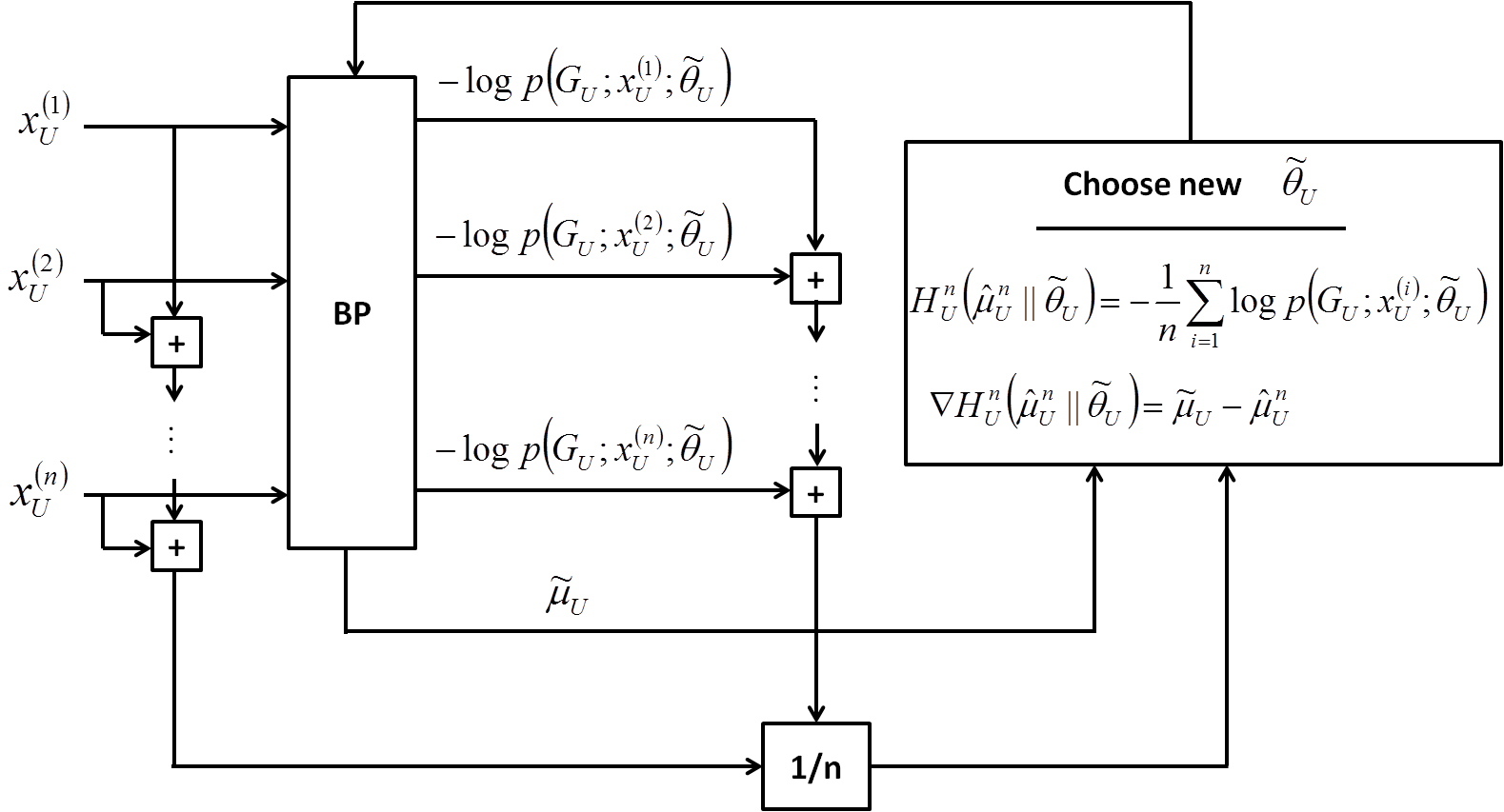}
    }   }
    \caption{ Block diagram for finding the moment-matching parameter $\theta^*_U$ for encoding $X_U$.}
    \label{fig:block_moment}
\end{figure*}

\vspace{2mm}

\begin{proposition}\label{prop:emp_cross}
    \begin{eqnarray}
        H^n_U(\hat\mu^n_U || \tilde\theta_U) & = & \Phi_U(\tilde\theta_U) - \langle \hat\mu^n_U,\tilde\theta_U\rangle \nonumber \\[2ex]
        \nabla H_U^n(\hat\mu^n_U || \tilde\theta_U) & = & \tilde\mu_U - \hat\mu^n_U \nonumber \\[1ex]                                                    & = & \tilde\Lambda_U(\tilde\theta_U) - \hat\mu^n_U ,\nonumber
    \end{eqnarray}
    where the gradient is with respect to $\tilde \theta_U$.
\end{proposition}

\vspace{2mm}

\begin{proof}
    Using relation (\ref{eq:mrf_1}) for the reduced MRF on $G_U$ with parameter $\tilde\theta_U$, we get
    \begin{eqnarray}
        H^n_U(\hat\mu^n_U || \tilde\theta_U)\!\!\!\! & = & \!\!\!\! \frac{1}{n}\sum\limits_{i=1}^n\left[\Phi_U(\tilde\theta_U)-\big \langle t(\bfx^{(i)}_U),\tilde\theta_U \big \rangle\right]\nonumber\\
                                                    & = & \Phi_U(\tilde\theta_U) - \Big \langle \sum\limits_{i=1}^nt(\bfx^{(i)}_U),\tilde\theta_U \Big \rangle\nonumber\\
                                                    & = & \Phi_U(\tilde\theta_U) - \big \langle \hat\mu^n_U,\tilde\theta_U \big \rangle\nonumber
    \end{eqnarray}
    It is well-known that $\nabla \Phi_U(\tilde\theta_U) = \tilde\mu_U$ \cite{wain:03b}. Then, taking the gradient of $H^n_U(\hat\mu^n_U || \tilde\theta_U)$ yields
    \begin{eqnarray}
        \nabla H^n_U(\hat \mu^n_U || \tilde\theta_U) & = & \nabla \frac{1}{n}\sum\limits_{i=1}^n\left[\Phi_U(\tilde\theta_U)- \big \langle t_U(\bfx^{(i)}_U),\tilde\theta_U \big\rangle \right] \nonumber \\
                                                     & = & \nabla \Phi_U(\tilde\theta_U) - \nabla \frac{1}{n}\sum\limits_{i=1}^n \big \langle t_U(\bfx^{(i)}_U),\tilde\theta_U \big \rangle \nonumber \\
                                                     & = & \nabla \Phi_U(\tilde\theta_U) - \nabla \Big \langle \frac{1}{n}\sum\limits_{i=1}^nt_U(\bfx^{(i)}_U),\tilde\theta_U \Big \rangle \nonumber \\
                                                     & = & \tilde\mu_U - \frac{1}{n}\sum\limits_{i=1}^nt_U(\bfx^{(i)}_U) \nonumber \\
                                                     & = & \tilde\mu_U - \hat\mu^n_U . \nonumber
    \end{eqnarray}
This completes the proof. $\hfill \Box$
\end{proof}

\vspace{2mm}

We now describe how a gradient descent algorithm can be used to find an estimate $\hat \theta^n_U$ of $\theta^*_U$ at which the gradient of $H^n_U(\hat \mu^n_U || \tilde\theta_U) $ is arbitrarily small. From the sequence $\bfx^{(1)}_U,\ldots,\bfx^{(n)}_U$, we first compute the empirical moment $\hat\mu^n_U = {1 \over n}\sum_{i=1}^nt_U(\bfx^{(i)}_U)$. Then, given a candidate parameter $\tilde\theta_{U}$, use Belief Propagation to compute the negative log-likelihood $-\log p(G_{U};\bfx^{(i)}_{U};\tilde\theta_{U})$ of the configuration $\bfx^{(i)}$ under the reduced MRF $p(G_{U};X_{U};\tilde\theta_{U})$, for each $i=1,\ldots,n$. Additionally, we compute the moment $\tilde\mu_{U}$ of the reduced MRF induced by the candidate parameter $\tilde\theta_U$, which like the probabilities, can be computed due to tractability of $U$. We then compute the objective function $H^n_{U}(\hat\mu^n_U || \tilde\theta_U) = - {1 \over n} \sum_{i=1}^n\log p(G_U;\bfx^{(i)}_U;\tilde\theta_U)$ and the gradient $\nabla H^n_U(\hat\mu^n_U || \tilde\theta_U) = \tilde\mu_U - \hat\mu^n_U$. Finally, given a tolerance $\epsilon_{\mu}$, if  $\| \nabla H^n_U(\hat\mu^n_U || \tilde\theta_U) \| < \epsilon_{\mu}$, the algorithm terminates and we set $\hat\theta^n_U = \tilde\theta_U$ which corresponds to the estimated moment $\hat{\hat \mu}^n_U = \tilde\Lambda_U(\hat\theta^n_U)$ at which the algorithm is terminated. Note that by Proposition \ref{prop:emp_cross}, the estimated moment $\hat{\hat \mu}^n_U$ is within $\epsilon_{\mu}$ of $\hat\mu^n_U$. If $\| \nabla H^n_U(\hat\mu^n_U || \tilde\theta_U) \| \geq \epsilon_{\mu}$, we determine a new candidate parameter $\tilde\theta_U$ using a standard gradient descent method \cite{boyd2004} and repeat the above steps. This is illustrated in Figure \ref{fig:block_moment}.

\vspace{2mm}

\begin{proposition}
    The estimate $\hat\theta^n_U$ is consistent, i.e.,
    for any $\epsilon >0$,
 \begin{eqnarray}  \label{eq:convergeinprob}
     \Pr \big( \big| \hat \theta^n_U - \theta^*_U \big| \leq \epsilon \big) \rightarrow 1, \mbox{ as } n \rightarrow \infty .
\end{eqnarray}
\end{proposition}

\vspace{2mm}

\begin{proof}
    Let $B(\theta^*_U,\epsilon_{\theta})$ be the $\epsilon_{\theta}$-ball centered at $\theta^*_U$. Assume without loss of generality that $B(\theta^*_U,\epsilon_{\theta})\subset\tilde\Theta_U$. Then, let $\epsilon_{\mu}$ be the largest tolerance around $\mu_U$ such that the $\epsilon_{\mu}$-ball $B(\mu_U,\epsilon_{\mu})$ centered at $\mu_U$ is contained in $\tilde \Lambda_U(B(\theta^*_U,\epsilon_{\theta}))$. It follows that
    \begin{eqnarray}
        \Pr \big( \big| \hat \theta^n_U - \theta^*_U \big| \leq \epsilon_{\theta} \big) & = & \Pr \big( \hat{ \hat \mu}^n_U \in \tilde \Lambda_U(B(\theta^*_U,\epsilon_{\theta})) \big) \nonumber \\
                & \geq & \Pr \big( \hat{\hat \mu}^n_U \in B(\mu_U,\epsilon_{\mu}) \big) \nonumber \\
                & = & \Pr \big( \big| \hat{\hat \mu}^n_U - \mu_U \big| \leq \epsilon_{\mu} \big) . \nonumber
    \end{eqnarray}

    Now let $\epsilon'_{\mu}=\epsilon_{\mu}/2$ be the tolerance on $\|\nabla H^n_U(\hat\mu^n_U || \tilde\theta_U)\|$ in the gradient descent algorithm. This means that $| \hat{\hat \mu}^n_U - \hat\mu^n_U | \leq \epsilon'_{\mu}$, which in turn implies that
    \begin{eqnarray}
        \Pr \big( \big| \hat{\hat \mu}^n_U - \mu_U \big| \leq \epsilon_{\mu} \big) & = & \Pr \big( \big| \hat \mu^n_U - \mu_U \big| \leq \epsilon'_{\mu} \big) . \nonumber
    \end{eqnarray}

    Using Proposition \ref{prop:momconv}, we can now say that for an arbitrary tolerance $\delta > 0$, there exists $N$ such that if the number of observations $n$ is greater than or equal to $N$, then
    \begin{eqnarray}
        \Pr \big( \big| \hat \theta^n_U - \theta^*_U \big| \leq \epsilon_{\theta} \big) & \geq & \Pr \big( \big| \hat \mu^n_U - \mu_U \big| \leq \epsilon'_{\mu} \big) \nonumber \\
                & \geq & 1 - \delta . \nonumber
    \end{eqnarray}

    This completes the proof. $\hfill \Box$
\end{proof}

\vspace{2mm}

\section{Tradeoffs between Lines and Strips}\label{sec:tradeoffs}

The following proposition shows that, as intuited earlier,  strip rate increases with strip width.

\vspace{2mm}

\begin{proposition}\label{prop:strips}
    \begin{eqnarray}
        \bar R^S_{n+1} & > & \bar R^S_n .\nonumber
    \end{eqnarray}
\end{proposition}

\vspace{2mm}

\begin{lemma}\label{lemma:firstRow_strip}
    Let $r_1$ denote the first row of rectangular region $B_n$ of sites of height $n$. Then,
    \begin{eqnarray}
        H(G;X_{r_1} | X_{\partial B_n};\theta) & < & H(G;X_{r_1} | X_{\partial B_{n+1}};\theta) .
    \end{eqnarray}
\end{lemma}

\vspace{2mm}

\begin{proof}
Note    $B_{n+1}$ consists of $B_n$ and an additional row $r_{n+1}$, which is part of the boundary of $B_n$. By the Markov property, $H(G;X_{r_1} | X_{\partial B_n};\theta) = H(G;X_{r_1} | X_{\partial B_{n+1}},X_{r_{n+1}};\theta)$. That is, conditioning on $\partial B_{n+1}$ and $r_{n+1}$ is the same as conditioning on $\partial B_n$. Finally, $H(G;X_{r_1} | X_{\partial B_{n+1}},X_{r_{n+1}};\theta) < H(G;X_{r_1} | X_{\partial B_{n+1}};\theta)$ as the left side has more conditioning. In summary
    \begin{eqnarray}
        H(G;X_{r_1} | X_{\partial B_n};\theta) & = & H(G;X_{r_1} | X_{\partial B_{n+1}},X_{r_{n+1}};\theta)\nonumber\\
                                        & < & H(G;X_{r_1} | X_{\partial B_{n+1}};\theta)\nonumber.
    \end{eqnarray}
    This completes the proof of Lemma \ref{lemma:firstRow_strip}. $\hfill \Box$
\end{proof}

\vspace{2mm}

\noindent We continue with the proof of Proposition \ref{prop:strips}.

\vspace{2mm}

\begin{proof}
    By direct calculation we have for a strip of height $n+1$ that
    \begin{eqnarray}
       \bar R^S_{n+1} & = & \frac{1}{(n+1)}H(G;X_{B_{n+1}} | X_{\partial B_{n+1}};\theta)\nonumber\\
                  & = & \frac{1}{(n+1)}H(G;X_{B_n} | X_{\partial B_n};\theta)
           + \frac{1}{(n+1) }H(G;X_{r_1} | X_{\partial B_{n+1}};\theta) , \label{eq:R_S_n+1}
    \end{eqnarray}
    and for a strip of height $n$,
    \begin{eqnarray}
        \bar R^S_n & = & \frac{1}{n}H(G;X_{B_n} | X_{\partial B_n};\theta)\nonumber\\
              & = & \frac{n+1}{n}\frac{1}{(n+1)}H(G;X_{B_n} | X_{\partial B_n};\theta)\nonumber\\
              & = & \frac{1}{(n+1)}H(G;X_{B_n} | X_{\partial B_n};\theta)
              + \frac{1}{n(n+1)}H(G;X_{B_n} | X_{\partial B_n};\theta)\nonumber\\
              & = & \frac{1}{(n+1)}H(G;X_{B_n} | X_{\partial B_n};\theta)
                + \frac{1}{n}\sum\limits_{i=1}^n\frac{1}{(n+1)}H(G;X_{r_i} | X_{\partial B_{n-i+1}};\theta)\nonumber\\
              & < & \frac{1}{(n+1)}H(G;X_{B_n} | X_{\partial B_n};\theta)
         + \frac{1}{(n+1)}H(G;X_{r_1} | X_{\partial B_{n+1}};\theta)\nonumber\\
              & = & \bar R^S_{n+1}\nonumber
    \end{eqnarray}
    by (\ref{eq:R_S_n+1}) and Lemma \ref{lemma:firstRow_strip}. This completes the proof. $\hfill \Box$
\end{proof}

\vspace{2mm}


Likewise, the next proposition shows that, as supposed earlier, line rate decreases with line width.

\vspace{2mm}

\begin{proposition}\label{prop:lines}
    \begin{eqnarray}
        \bar R^L_{n+1} & < & \bar R^L_n . \nonumber
    \end{eqnarray}
\end{proposition}

\vspace{2mm}

\begin{proof}
First we note that reducing $X_{B_{n+1}}$ to $\tilde X_{B_{n+1}}$ by matching
moments and further reducing the $X_{B_n}$ marginal of  $\tilde X_{B_{n+1}}$
to $\tilde X_{B_n}$ by matching moments results in the same reduced MRF
on $G_{B_n}$ as would reducing the original $X_{B_n}$ to $\tilde X_{B_n}$
by matching moments.
Let $\theta^*_n$ be the moment matching parameter for $\tilde X_{B_n}$.
    \begin{eqnarray}
        \bar R^l_{n+1} & = & \frac{1}{n+1}H(G_{B_{n+1}};X_{B_{n+1}};\theta^*_{n+1})\nonumber\\
                  & = & \frac{1}{n+1}\left[H(G_{B_{n+1}};X_{B_{n}};\theta^*_{n+1})  +   H(G_{B_{n+1}};X_{r_{n+1}}|X_{B_n};\theta^*_{n+1})\right]\nonumber\\
                  & < & \frac{1}{n+1}\left[H(G_{B_{n+1}};X_{B_{n}};\theta^*_{n+1})  + \frac{1}{n}H(G_{B_{n+1}};X_{B_{n}};\theta^*_{n+1})\right]\nonumber\\
                  & = & \frac{1}{n}H(G_{B_{n+1}};X_{B_{n}};\theta^*_{n+1})\nonumber\\
                  & < & \frac{1}{n}H(G_{B_{n}};X_{B_{n}};\theta^*_{n})\nonumber\\
                  & = & \bar R^L_n ,  \nonumber
    \end{eqnarray}
    \noindent where the second inequality is from the maximum entropy property of MRFs. This completes the proof $\hfill \Box$
\end{proof}

\vspace{2mm}


\vspace{2mm}

\begin{proposition}\label{prop:lines_over_strips}
    For all strip widths $n_S$ and line widths $n_L$,
    \begin{eqnarray}
        \bar R^L_{n_L} & > & \bar R^S_{n_S} . \nonumber
    \end{eqnarray}
\end{proposition}

\vspace{2mm}

\begin{proof}
    We prove the proposition by cases: $n_S = n_L$, $n_S > n_L$, and $n_S < n_L$.

    First assume $n_S=n_L=n$. Then,
    \begin{eqnarray}
       \bar R^S_{n_S} & = & \frac{1}{n}H(G;X_{B_n}|X_{\partial B_n};\theta) \nonumber \\
                  & \leq & \frac{1}{n}H(G;X_{B_n};\theta) \nonumber \\
                  & < & \frac{1}{n}H(G_{B_n};X_{B_n};\theta^*_n) \label{eq:max1} \\
                  & = & \bar R^L_{n_L} , \nonumber
    \end{eqnarray}
    where (\ref{eq:max1}) follows from the maximum entropy property of MRFs.
    Next, assume $n_S>n_L$. Then,
    \begin{eqnarray}
       \bar R^S_{n_S} & = & \frac{1}{n_S}H(G;X_{B_n}|X_{\partial B_n};\theta) \nonumber \\
                  & \leq & \frac{1}{n_S}H(G;X_{B_n};\theta) \nonumber \\
                  & < & \frac{1}{n_S}H(G_{B_n};X_{B_{n_S}};\theta^*_{n_S}) \label{eq:max2} \\
                  & = & \bar R^L_{n_S} \nonumber \\
                  & < & \bar R^L_{n_L} ,  \nonumber
    \end{eqnarray}
    where (\ref{eq:max2}) follows from the maximum entropy property of MRFs.
    Finally, assume $n_S<n_L$. Then,
    \begin{eqnarray}
        \bar R^S_{n_S} & < & \bar R^S_{n_L} \nonumber \\
                  & = & \frac{1}{n_L}H(G;X_{B_{n_L}}|X_{\partial B_{n_L}};\theta) \nonumber \\
                  & \leq & \frac{1}{n_L}H(G;X_{B_{n_L}};\theta) \nonumber \\
                  & < & \frac{1}{n_L}H(G_{B_{n_L}};X_{B_{n_L}};\theta^*_{n_L}) \label{eq:max3} \\
                  & = &\bar  R^L_{n_L} , \nonumber
    \end{eqnarray}
    where (\ref{eq:max3}) follows from the maximum entropy property of MRFs.
This completes the proof.  $\hfill  \Box$
\end{proof}

\begin{figure*}
    \centerline{    \hbox{
    \hspace{0.0in}
    \includegraphics[scale = .35]{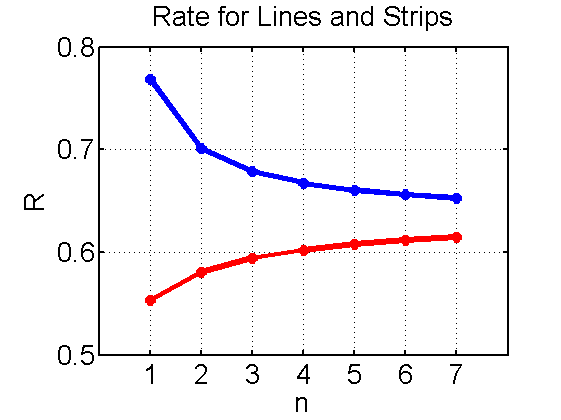}
    \hspace{1.0in}
    \includegraphics[scale = .35]{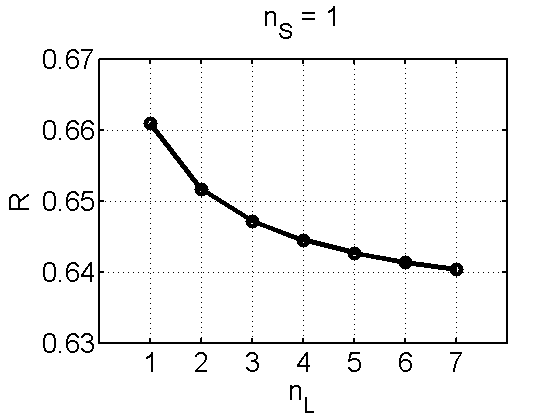}
    }   }
    \hbox{\small \hspace{1.38in} (a) \hspace{2.9in} (b)}
    \vspace{3mm}
    \centerline{    \hbox{
    \hspace{0.0in}
    \includegraphics[scale = .35]{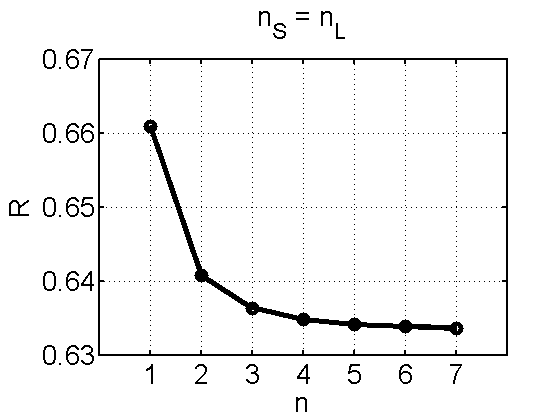}
    \hspace{1.0in}
    \includegraphics[scale = .35]{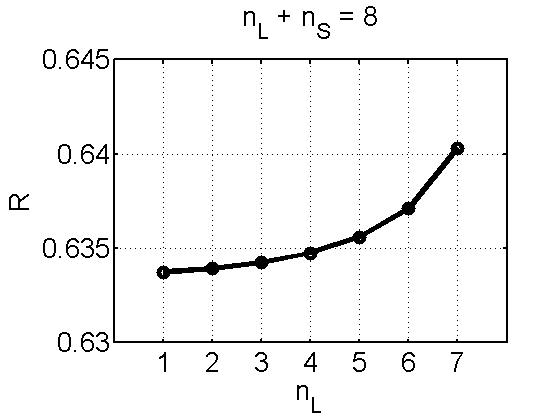}
    }   }
    \hbox{\small \hspace{1.38in} (c) \hspace{2.9in} (d)}
    \caption{ Rate (a) for lines (blue) and strips (red); (b) as a function of $n_L$ for $n_S=1$; (c) as a function of $n=n_S=n_L$; and (d) for $n_S+n_L = 8$.}
    \label{fig:example}
\end{figure*}

Together these three propositions indicate that $\bar R^L_{n_L}$ and $\bar R^S_{n_S}$ always behave as in Figure \ref{fig:example} (a), which as discussed in the next section, plots them for a specific case.  They also illustrate the potential tradeoffs between line width $n_L$ and strip width $n_S$. Specifically, by increasing $n_L$ the line rate $\bar R^L_{n_L}$ decreases, though the fraction $\frac{n_L}{n_S+n_L}$ of pixels encoded at the higher rate increases, while increasing $n_S$ increases the fraction $\frac{n_S}{n_L+n_S}$ of pixels encoded at the lower rate, though the strip rate $\bar R^S_{n_S}$ increases.

In addition to considering the effect of $n_S$ and $n_L$ on rate, we can look at their influence on the rate redundancy $\Delta (n_S, n_L) \defeq \frac{1}{|V|}D(X_U || \tilde X_U)$, which is entirely due to encoding the lines independently and as moment-matching reduced MRFs. We use the shorthand notation $\tilde X_{B_{n_L}}$ to indicate the moment-matching reduced MRF on $B_{n_L}$ and $D(X_{B_{n_L}}||\tilde X_{B_{n_L}})$ to denote the divergence between the marginal and moment-matching reduced MRF distributions for $X_{B_{n_L}}$.

\vspace{2mm}

\begin{proposition}\label{prop:redundancy}
    The per-row rate redundancy due to coding  $X_U\sim p(G;X_U;\theta)$ as a reduced MRF $X_U\sim p(G_U;X_U:\theta^*_U)$ is
\begin{eqnarray}
  \bar \Delta(n_S, n_L) &=&  {n_S \over n_S + n_L} I(X_{r_1};X_{r_{-n_S}}) + {n_L \over n_S + n_L} D(X_{B_{n_L}}||\tilde X_{B_{n_L}}) ,  \nonumber
\end{eqnarray}
    where
    $r_1$ is the 1st row of a line, and $r_{-n_S}$ is the last row of the previous line.
\end{proposition}

\vspace{2mm}

\begin{proof}
    To prove the proposition, consider a joint distribution $p(x_1,\ldots,x_N)$ on $N$ variables, where we have in mind each variable representing one of the $N=k+1$ lines. By approximating $p(x_1,\ldots,x_N)$ with $\tilde p(x_1,\ldots,x_N)=\prod_{i=1}^N\tilde p(x_i)$ we can see that the divergence between $p$ and $\tilde p$ is
    \begin{eqnarray}
        D(p || \tilde p) & = & \sum\limits_{x_1,\ldots,x_N}p(x_1,\ldots,x_N)\log\frac{p(x_1,\ldots,x_N)}{\tilde p(x_1)\cdots\tilde p(x_N)} \nonumber \\
                         & = & -\sum\limits_{x_1,\ldots,x_N}p(x_1,\ldots,x_N)\log\tilde p(x_1)\cdots\tilde p(x_N) - H(X_1,\ldots,X_N) \nonumber \\
                         & = & \sum\limits_{i=1}^N\sum\limits_{x_i}-p(x_i)\log\tilde p(x_i) - H(X_1,\ldots,X_N) \nonumber \\
                         & = & \sum\limits_{i=1}^N\left[H(X_i) + D(p(X_i)||\tilde p(X_i))\right] - H(X_1,\ldots,X_N) \nonumber \\
                         & = & \sum\limits_{i=1}^N\left[H(X_i) - H(X_i|X_{i-1},\ldots,X_1) + D(p(X_i)||\tilde p(X_i))\right] \nonumber \\
                         & = & \sum\limits_{i=2}^N I(X_i;X_{i-1}) + \sum\limits_{i=1}^N D(p(X_i) || \tilde p(X_i)) . \nonumber
    \end{eqnarray}
    Applying the stationarity assumption, weighting the last two terms by the (approximate) fractions in (\ref{eq:rate_approx}), and substituting $N=k+1$ and $X_i=X_{B_{n_L}}$ yields
    \begin{eqnarray}
        \bar \Delta(n_S, n_L) & = & {n_S \over n_S + n_L} I(X_{B_{n_L}};X_{B_{n_L},-n_S}) + {n_L \over n_S + n_L} D(X_{B_{n_L}}||\tilde X_{B_{n_L}}) , \nonumber
    \end{eqnarray}
    where $I(X_{B_{n_L}};X_{B_{n_L},-n_S})$ is the mutual information between two $n_L\times N$ rectangular blocks of sites separated by a $n_S\times N$ rectangular block of sites. To finish the proof, it suffices to consider $I(X_1,X_2;Y_1,Y_2)$ where $X_1-X_2-Y_1-Y_2$ form a Markov Chain. In this case,
    \begin{eqnarray}
        I(X_1,X_2;Y_1,Y_2) & = & H(Y_1,Y_2) - H(Y_1,Y_2 | X_1,X_2) \nonumber \\
                           & = & H(Y_1) + H(Y_2 | Y_1) - H(Y_1 | X_1,X_2) - H(Y_2 | Y_1,X_1,X_2) \nonumber \\
                           & = & H(Y_1) + H(Y_2 | Y_1) - H(Y_1 | X_2) - H(Y_2 | Y_1) \nonumber \\
                           & = & H(Y_1) - H(Y_1 | X_2) \nonumber \\
                           & = & I(Y_1;X_2)  . \nonumber
    \end{eqnarray}
    Making the appropriate substitutions yields
    \begin{eqnarray}
        \bar \Delta(n_S, n_L) & = & {n_S \over n_S + n_L} I(X_{r_1};X_{r_{-n_S}}) + {n_L \over n_S + n_L} D(X_{B_{n_L}}||\tilde X_{B_{n_L}}) , \nonumber
    \end{eqnarray}
    where $I(X_{r_1};X_{r_{-n_S}})$ is the mutual information between the 1st row of a line and the last row of the previous line. This completes the proof.  $\hfill \Box$
\end{proof}

\vspace{2mm}

This proposition shows specifically how the redundancy of RCC has two components: a correlation redundancy $I(X_{r_1};X_{r_{-n_S}}) $ due to encoding the lines independently of one another, and a distribution redundancy $D(X_{B_{n_L}}||\tilde X_{B_{n_L}})$ due to approximating the lines as moment matching reduced MRFs.

\vspace{2mm}

\begin{proposition}\label{prop:info}
    $I(X_{r_1};X_{r_{-n_S}})$ is decreasing in $n_S$.
\end{proposition}

\vspace{1mm}

\begin{proof}
    We let $X_{r_{i,1}}$ denote the 1st row of the $i$-th line and $X_{r_{i-1,n_L}}$ and $X_{r_{i-1,n_L-1}}$ denote, respectively, the $n_L$-th and $(n_L-1)$-st lines of the $(i-1)$-st line.
    \begin{eqnarray}
        I(X_{r_1};X_{r_{-n_S}}) & = & H(G;r_{i,1};\theta) - H(G;X_{r_{i,1}} | X_{r_{i-1,n_L}};\theta) \nonumber \\
                                & = & H(G;r_{i,1};\theta) - H(G;X_{r_{i,1}} | X_{r_{i-1,n_L}}, X_{r_{i-1,n_L-1}};\theta) \label{eq:mark} \\
                                & > & H(G;r_{i,1};\theta) - H(G;X_{r_{i,1}} | X_{r_{i-1,n_L-1}};\theta) \label{eq:condineq} \\
                                & = & I(X_{r_1};X_{r_{-(n_S+1)}}) , \nonumber
    \end{eqnarray}
    where (\ref{eq:mark}) is due to the Markov property and (\ref{eq:condineq}) is due to removing conditioning. This completes the proof. $\hfill \Box$
\end{proof}

\vspace{2mm}


To analyze the distribution redundancy, we let $\tilde{\tilde {X}}_{B_{n}}$ be the marginal distribution of $X_{B_{n-1}}$ as a subset of the moment-matching reduced MRF  ${\tilde {X}}_{B_{n}}$on $B_{n}$. More generally, $X_{B_n}$ decorated with $k$ ``tildes" indicates the marginal distribution of $X_{B_{n-k+1}}$ as a subset of the moment-matching reduced MRF $\tilde X_{B_{n}}$ on $B_{n}$.  Moreover, we let $\theta^*_n$ be shorthand for $\theta^*_{B_{n}}$. We then have the following recursive expression for the distribution redundancy.

\vspace{2mm}

\begin{proposition}\label{conj:div}
    \begin{eqnarray}
       D(X_{B_{n_L}}|| \tilde X_{B_{n_L}})
             &   = & \! D(X_{B_{n_L-1}}||\tilde X_{B_{n_L-1}})
                                          - D(\tilde{\tilde{X}}_{B_{n_L}}||\tilde X_{B_{n_L-1}})\nonumber\\[.4ex]
              & &  + \, H(G_{B_{n_L}};r_{n_L}|r_{n_L-1};\theta^*_{n_L})
                                          - H(G;r_{n_L}|r_{n_L-1};\theta) , \nonumber
    \end{eqnarray}
    where $D(\tilde{\tilde X}_{B_{n_L}}||\tilde X_{B_{n_L-1}})$ is the divergence between the marginal distribution of $X_{B_{n_L-1}}$ as a subfield of $\tilde X_{B_{n_L}}$ and the reduced MRF $\tilde X_{B_{n_L-1}}$ on $B_{n_L-1}$, and where $H(\cdot ; r_n | r_{n-1} ; \cdot)$ is the conditional
    entropy of row $r_n$ condition on row $r_{n-1}$ for the specified graph and parameter vector.
\end{proposition}

\vspace{2mm}

\begin{proof}
    We prove the proposition by using the fact that the divergence $D(X_{B_{n_L}}|| \tilde X_{B_{n_L}})$ between the marginal distribution of $X_{B_{n_L}}$ and the reduced MRF for $X_{B_{n_L}}$ can be expressed as the difference between the entropy of the latter and that of the former. Specifically,
    \begin{eqnarray}
        D(X_{B_{n_L}}|| \tilde X_{B_{n_L}}) & = & H(G_{B_{n_L}};X_{B_{n_L}};\theta^*_{n_L}) - H(G;X_{B_{n_L}};\theta) \nonumber \\
                                            & =  & H(G_{B_{n_L}};X_{B_{n_L-1}};\theta^*_{n_L}) - H(G;X_{B_{n_L-1}};\theta) \nonumber \\
                                            &   & + H(G_{B_{n_L}};r_{n_L} | r_{n_L-1};\theta^*_{n_L}) - H(G;r_{n_L} | r_{n_L-1};\theta) \nonumber \\
                                            & =  & H(G_{B_{n_L-1}};X_{B_{n_L-1}};\theta^*_{n_L-1}) - D(\tilde{\tilde{X}}_{B_{n_L}}|| \tilde X_{B_{n_L-1}}) \nonumber \\
                                            &    & - H(G;X_{B_{n_L-1}};\theta) + H(G_{B_{n_L}};r_{n_L} | r_{n_L-1};\theta^*_{n_L}) - H(G;r_{n_L} | r_{n_L-1};\theta) \nonumber \\
                                            & =  & H(G_{B_{n_L-1}};X_{B_{n_L-1}};\theta^*_{n_L-1}) - H(G;X_{B_{n_L-1}};\theta)  \nonumber \\
                                            &    & - D(\tilde{\tilde{X}}_{B_{n_L}}|| \tilde X_{B_{n_L-1}}) + H(G_{B_{n_L}};r_{n_L} | r_{n_L-1};\theta^*_{n_L}) - H(G;r_{n_L} | r_{n_L-1};\theta) \nonumber \\
                                            & =  & D(X_{B_{n_L-1}}|| \tilde X_{B_{n_L-1}}) - D(\tilde{\tilde{X}}_{B_{n_L}}|| \tilde X_{B_{n_L-1}}) \nonumber \\
                                            &    &  + H(G_{B_{n_L}};r_{n_L} | r_{n_L-1};\theta^*_{n_L}) - H(G;r_{n_L} | r_{n_L-1};\theta) . \nonumber
    \end{eqnarray}
    This completes the proof. $\hfill \Box$
\end{proof}

\vspace{2mm}

Furthermore, the divergence $D(\tilde{\tilde{X}}_{B_{n_L-1}}||\tilde X_{B_{n_L-1}})$
has the following recursive relationship.

\vspace{2mm}

\begin{proposition}
\begin{eqnarray}
    D(\tilde{\tilde{X}}_{B_{n-k+1}}||\tilde X_{B_{n-k}})  & = & D(\tilde{\tilde{\tilde{X}}}_{B_{n-k+1}}||\tilde X_{B_{n-k-1}}) -   D(\tilde{\tilde{X}}_{B_{n-k}}||\tilde X_{B_{n-k-1}})\nonumber\\[.5ex]
             &&    + H(G_{B_{n-k}};r_{n-k}|r_{n-k-1};\theta^*_{n-k}) - H(G_{B_{n-k+1}};r_{n-k}|r_{n-k-1};\theta^*_{n-k+1})\nonumber
\end{eqnarray}
where $D(\tilde{\tilde{\tilde{X}}}_{B_{n-k+1}}||\tilde X_{B_{n-k-1}})$ is the divergence between the marginal distribution of $X_{B_{n-k-1}}$ as a subfield of $\tilde X_{B_{n-k+1}}$ and the reduced MRF $\tilde X_{B_{n-k-1}}$ on $B_{n-k-1}$.
\end{proposition}

\vspace{2mm}

\begin{proof}
    \begin{eqnarray}
        D(\tilde{\tilde{X}}_{B_{n-k+1}}||\tilde X_{B_{n-k}}) & = & H(G_{B_{n-k}};X_{B_{n-k}};\theta^*_{n-k}) - H(G_{B_{n-k+1}};X_{B_{n-k}};\theta^*_{n-k+1}) \nonumber \\
                                                             & = & H(G_{B_{n-k}};X_{B_{n-k-1}};\theta^*_{n-k}) - H(G_{B_{n-k+1}};X_{B_{n-k-1}};\theta^*_{n-k+1})  \nonumber \\
                                                             &   & + H(G_{B_{n-k}};r_{n-k} | r_{n-k-1};\theta^*_{n-k}) - H(G_{B_{n-k+1}};r_{n-k} | r_{n-k-1};\theta^*_{n-k+1}) \nonumber \\
                                                             & = & H(G_{B_{n-k-1}};X_{B_{n-k-1}};\theta^*_{n-k-1}) - D(\tilde{\tilde{X}}_{n-k} || \tilde{X}_{n-k-1}) \nonumber \\
                                                             &   & - H(G_{B_{n-k+1}};X_{B_{n-k-1}};\theta^*_{n-k+1})  + H(G_{B_{n-k}};r_{n-k} | r_{n-k-1};\theta^*_{n-k}) \nonumber \\
                                                             &   & - H(G_{B_{n-k+1}};r_{n-k} | r_{n-k-1};\theta^*_{n-k+1}) \nonumber \\
                                                             & = & D(\tilde{\tilde{\tilde{X}}}_{n-k+1} || \tilde{X}_{n-k-1}) - D(\tilde{\tilde{X}}_{n-k} || \tilde{X}_{n-k-1}) \nonumber \\
                                                             &   & + H(G_{B_{n-k}};r_{n-k} | r_{n-k-1};\theta^*_{n-k}) - H(G_{B_{n-k+1}};r_{n-k} | r_{n-k-1};\theta^*_{n-k+1}) . \nonumber
        \end{eqnarray}
        This completes the proof. $\hfill \Box$
\end{proof}

\vspace{2mm}

Intuitively we would expect the term
$D(X_{B_{n_L}}||\tilde X_{B_{n_L}})$ to decrease in $n_L$, as this divergence is zero when $n_L = M$ and indeed we conjecture that this is the case.  At the very least, we
expect ${1 \over n_L} D(X_{B_{n_L}}||\tilde X_{B_{n_L}})$ to decrease in $n_L$.

We now consider the effects of changing $n_S$ and $n_L$ on redundancy,
as expressed in Proposition \ref{prop:redundancy}.
Increasing $n_S$ decreases distribution redundancy through the factor ${n_L \over n_S + n_L}$.
It is not so clear what happens to the correlation redundancy, as increasing $n_S$
increases the fraction ${n_S \over n_S + n_L}$, while decreasing the information
$I(X_{r_1};X_{r_{-n_S}})$. However, if we keep $n_S$ and $n_L$ proportional to one another, as $n_S$ increases,
the fraction stays the same, the correlation redundancy decreases, and assuming the
conjecture, so too does distribution redundancy.

Similarly, increasing $n_L$ decreases the correlation redundancy through the factor
${n_S \over n_S + n_L}$.  Even assuming the above conjecture, it is
not  clear what happens to the distribution redundancy, as increasing $n_L$
increases the fraction ${n_L \over n_S + n_L}$, while decreasing the divergence
$D(X_{B_{n_L}}||\tilde X_{B_{n_L}})$. However, as mentioned above, if $n_S$ and $n_L$ increase proportionally to one another, then the fraction stays the same and both the correlation and distribution redundancies decrease in $n_L$.



The complexity of this coding scheme can be expressed as
\begin{eqnarray}
    C_{n_S,n_L} & = & \frac{n_S}{n_S+n_L}|\mathcal{X}|^{n_S}c_S + \frac{n_L}{n_S+n_L}|\mathcal{X}|^{n_L}c_L  \nonumber
\end{eqnarray}
\noindent where $| \mathcal{X}|$ denotes the number of elements of $\mathcal X$, and $c_S$ and $c_L$ are factors relating the complexity of encoding a strip versus a line. For example, numerical simulations show that for $n_S=n_L$, the run-time involved in encoding a strip is a little higher than that for a line, which is due to additional operations for conditioning on the boundary of a strip. However, the difference
becomes negligible as $n_S$ and $n_L$ become larger.  As a result, the complexity $C_{n_S,n_L}$ is dominated by $\max\{n_S,n_L\}$. Given a constraint $\max\{n_S,n_L\}\leq n^*$ on the maximum exponent in the complexity, since both Proposition \ref{prop:info} and our conjecture indicate choosing $n_S$ and $n_L$ each to be as large as possible, we propose setting $n_S = n_L$.

\section{Example: Homogeneous Ising Model}\label{sec:simulation}

We simulated a homogeneous Ising model with edge parameter $\theta_{ij}=0.4$ and node parameter $\theta_i = 0$ using Gibbs sampling. To encode the lines with line width $n_L$, we approximate the moment-matching parameter $\theta^*_{n_L}$ by minimizing the empirical cross entropy
\begin{eqnarray}
    H^{nK}_{n_L}(\tilde\theta_{n_L}) & = & \frac{1}{nK}\sum\limits_{L_i}\sum\limits_{j=1}^n -\log p(G_{L_i};\bfx^{(j)}_{L_i}; \tilde\theta_{n_L}).  \nonumber
\end{eqnarray}
\noindent Note that even for a homogeneous MRF, the moment-matching parameter for a subset $U$ will in general not be homogeneous.

The line rate $\bar R^L_{n_L}$ is approximated by
\begin{eqnarray}
    \hat R^L_{n_L} & = & \frac{1}{nK}\sum\limits_{L_i}\sum\limits_{j=1}^n -\log p(G_{L_i};\bfx^{(j)}_{L_i};\theta^*_{n_L})  . \nonumber
\end{eqnarray}

Similarly, $\bar R^S_{n_S}$ is approximated by
\begin{eqnarray}
    \hat R^S_{n_S} & = & \frac{1}{nK}\sum\limits_{S_i}\sum\limits_{j=1}^n -\log p(G;\bfx^{(j)}_{S_i}|\bfx^{(j)}_{\partial S_i};\theta) . \nonumber
\end{eqnarray}

Figure \ref{fig:example}(a) shows $\hat R^L_{n_L}$ and $\hat R^L_{n_S}$.
As predicted by Propositions \ref{prop:strips}, \ref{prop:lines}, and \ref{prop:lines_over_strips}, $\hat R^S_{n_S}$ is increasing in $n_S$, $\hat R^L_{n_L}$ is decreasing in $n_L$, and $\hat R^S_{n_S} < \hat R^L_{n_L}$ for all $n_S,n_L$.
We computed $\hat R_{n_S,n_L}$ from $\hat R^L_{n_L}$ and $\hat R^L_{n_S}$
using (\ref{eq:RSL}), and as seen in Figure \ref{fig:example}(b), we
found that $\hat R_{n_S,n_L}$  decreases as $n_L$ increases for constant $n_S$.
We also found, see Figure \ref{fig:example}(c), that $\hat R_{n_S,n_L}$  decreases with $n$ increasing when $n =n_L = n_S$, which is consistent with the earlier discussion that presumed the conjecture.
Finally, we found that if one holds the sum $n_L +n_S$ constant, then the rate $\hat R_{n_S,n_L}$  is minimized
when $n_L = 1$.  This indicates that the information  $I(X_{r_1};X_{r_{-n_S}})$ decreases with $n_S$ faster than the
divergence $D(X_{B_{n_L}}||\tilde X_{B_{n_L}})$ decreases with $n_L$.
Though not apparent in the Figure, we found that $\hat R_{7,7} < \hat R_{7,1}$, an improvement over  our earlier paper \cite{reyes2010} which focused exclusively on  $n_L = 1$. However, the improvement is nominal, so therefore, at least for this particular value of $\theta_{ij}$, does not justify the significantly increased complexity.


%
%
%



\section{Concluding Remarks}
In this paper we have addressed the topic of tradeoffs in the choice of the width $n_L$ and spacing $n_S$ of the cutset components in Reduced Cutset Coding of Markov random fields. We have provided analysis from the perspective of the rate of this scheme in terms of the rates for encoding lines and strips and the relative contributions of each to the overall rate. We have shown that the rate for encoding lines with the moment-matching reduced MRF decreases with $n_L$, and that the rate for encoding strips increases with $n_S$, and on the basis of just these results one might conclude that large $n_L$ and small $n_S$ would provide an optimal combination. However, we also show that for all combinations of $n_L$ and $n_S$, the rate for encoding lines is strictly greater than the rate for encoding strips. Moreover, the fraction ${n_L \over n_S + n_L}$ of sites encoded at the larger rate obviously increases with $n_L$, while the fraction ${n_S \over n_S + n_L}$ of sites encoded at the smaller rate obviously decreases with $n_S$.

Additionally, we have analyzed the problem from the perspective of the redundancy in the code, showing that this redundancy decomposes into a distribution redundancy due to approximating the lines as moment-matching reduced MRFs, and a correlation redundancy due to independent coding of the lines. We show that the correlation redundancy is decreasing in $n_S$ and provide analysis of the distribution redundancy and conjecture that it is decreasing in $n_L$. Indeed, numerical experiments with an Ising model corroborate this conjecture. Moreover, if we let $n_L$ be the height of the original image, then clearly the divergence $D(X_{B_{n_L}} || \tilde X_{B_{n_L}})=0$, and at least offhand, there is no reason to suspect that this divergence is non-monotonic in $n_L$. Naturally, though, further analysis of $D(X_{B_{n_L}}||\tilde X_{B_{n_L}})$ remain to be done, and at least at the moment, we suspect that the recursive relations for $D(X_{B_{n_L}}||\tilde X_{B_{n_L}})$ will be useful in proving our conjecture.

While for general row-invariant statistics $t$ and exponential parameters $\theta$ it is not clear what the best choices of $n_L$ and $n_S$ should be, our numerical experiments with a uniform Ising model with parameters $\theta_{ij}=0.4,\theta_i=0$ suggest that letting $n_S$ and $n_L$ both be as large as possible achieves a lower rate. However, since the decrease in rate over a large $n_S$ and $n_L=1$ is in the fourth decimal place (in terms of per-site rate), the greatly increased complexity in encoding lines with large $n_L$ does not seem worth it. However, more work remains to be done in understanding how differences in parameter values affect these tradeoffs. And more generally, beyond the Ising model, we would like to understand how the apparent tradeoffs between $n_S$ and $n_L$ vary with $\theta$ for different types of statistic $t$. Previous work of the authors  \cite{reyes2009b,reyes2011,reyes2013} has looked at the relationship between {\em positively correlated} statistics $t$ and quantities of interest and it will be interesting to see if such statistics can be shown to have significant consequences for RCC.


\Section{References}

\end{document}